\documentclass[%
 reprint,
superscriptaddress,
 amsmath,amssymb,
 aps,
]{revtex4-2}

\usepackage{graphicx}
\usepackage{dcolumn}
\usepackage{bm}
\usepackage[plainpages=false, colorlinks=true, anchorcolor=blue, linkcolor=cyan, citecolor=cyan, bookmarks=false]{hyperref}
\usepackage{xcolor}
\usepackage{aas_macros}

\begin{document}

\preprint{APS/123-QED}

\title{A Parallelized Bayesian Approach To \\Accelerated Gravitational-Wave Background Characterization}

\author{Stephen~R.~Taylor}
    \affiliation{Department of Physics \& Astronomy, Vanderbilt University, 2301 Vanderbilt Place, Nashville, TN 37235, USA}
    \email{stephen.r.taylor@vanderbilt.edu}
\author{Joseph~Simon}
    \affiliation{Department of Astrophysical and Planetary Sciences, University of Colorado, Boulder, CO 80309, USA}
\author{Levi~Schult}
    \affiliation{Department of Physics \& Astronomy, Vanderbilt University, 2301 Vanderbilt Place, Nashville, TN 37235, USA}
    \affiliation{University of Virginia, 1827 University Avenue, Charlottesville, Virginia}
\author{Nihan~Pol}
    \affiliation{Department of Physics \& Astronomy, Vanderbilt University, 2301 Vanderbilt Place, Nashville, TN 37235, USA}
\author{William~G.~Lamb}
    \affiliation{Department of Physics \& Astronomy, Vanderbilt University, 2301 Vanderbilt Place, Nashville, TN 37235, USA}

\date{\today}

\begin{abstract}
The characterization of nanohertz-frequency gravitational waves (GWs) with pulsar-timing arrays requires a continual expansion of datasets and monitored pulsars. Whereas detection of the stochastic GW background is predicated on measuring a distinctive pattern of inter-pulsar correlations, characterizing the background's spectrum is driven by information encoded in the power spectra of the individual pulsars' time series. We propose a new technique for rapid Bayesian characterization of the stochastic GW background that is fully parallelized over pulsar datasets. This Factorized Likelihood (FL) technique empowers a modular approach to parameter estimation of the GW background, multi-stage model selection of a spectrally-common stochastic process and quadrupolar inter-pulsar correlations, and statistical cross-validation of measured signals between independent pulsar sub-arrays. We demonstrate the equivalence of this technique's efficacy with the full pulsar-timing array likelihood, yet at a fraction of the required time. Our technique is fast, easily implemented, and trivially allows for new data and pulsars to be combined with legacy datasets without re-analysis of the latter.
\end{abstract}

\maketitle


\section{Introduction} \label{sec:intro}

The precision timing of millisecond radio pulsars can be exploited over long observational campaigns to search for ultra low-frequency gravitational waves (GWs). As a GW deforms the spacetime between the Earth and a pulsar, it causes the times of arrival (TOAs) of radio pulses to deviate from flat spacetime expectations. These arrival-time deviations depend on the relative position of the GW source to the Earth--pulsar line-of-sight, and the GW metric perturbation at the time the wavefront first passed the pulsar as well as when it subsequently passes the Earth \citep{ew75,saz78,det79}. The result is that after detailed timing ephemerides are constructed for each pulsar---depending on its rotational period, period spindown, etc.---there will be an additional unmodeled GW-induced influence.

The GW frequencies to which pulsar-timing campaigns are sensitive depend on how the time-series of pulses is sampled; broadly speaking, the lower frequency is dictated by the inverse observation time (now greater than a decade, such that $f_\mathrm{low}\lesssim 3$~nHz), while the accessible upper frequencies are limited by the observational cadence and timing precision to $f_\mathrm{high}\sim 0.1$~$\mu\mathrm{Hz}$. The class of objects that are expected to dominate the GW emission in this band of frequencies are binary systems of supermassive black holes, with masses $\gtrsim 10^8-10^{10}M_\odot$ \citep{bbr80,wl03,jaf04,svc08,2019A&ARv..27....5B,2021arXiv210903262B}. These should be formed as a consequence of the hierarchical growth of galaxies over cosmic time, where resident supermassive black holes pair up within the common galactic merger remnant after a chain of dynamical processes. The GW signals from these supermassive black-holes binaries (SMBHBs) will not all be individually resolvable with pulsar timing, instead stacking incoherently within frequency-resolution bins to form a stochastic GW background (GWB) \citep{bs12,pbs+12,bp12}. 

In an individual pulsar's data, this GWB will manifest as a low-frequency stochastic (red) process, which could appear spectrally similar to intrinsic red noise from e.g., pulsar rotational instabilities, or pulse propagation through time-dependent ionized interstellar medium paths \citep{2010ApJ...725.1607S,2010arXiv1010.3785C,2020ApJ...892...89L,2016ApJ...821...66L}. Therefore, GW searches through pulsar timing are performed by correlating data across a pulsar-timing array (PTA) \citep{fb90}, where the sought-after evidence consists of a distinctive GWB-induced inter-pulsar correlation signature known as the Hellings \& Downs curve \citep{1983ApJ...265L..39H}. There are several collaborations that have now been contributing to this goal for over a decade, including the North American Nanohertz Observatory for Gravitational waves (NANOGrav) \citep{ml13}, the Parkes Pulsar Timing Array (PPTA) \citep{m08,mhb+13}, and the European Pulsar Timing Array (EPTA) \citep{kc13}. Together with the Indian Pulsar Timing Array (InPTA) \citep{inpta}, these collaborations also constitute the International Pulsar Timing Array (IPTA) \citep{vlh+16,pdd+19,2022MNRAS.tmp...73A}. The Chinese Pulsar Timing Array \citep{cpta}, and telescope-centered efforts such as CHIME/Pulsar \citep{chime} and MeerTime \citep{meertime}, will be important future contributors.

These PTA collaborations have recently delivered results that, for the first time, go beyond placing upper limits on the amplitude of a putative GWB. The NANOGrav Collaboration has announced the discovery a low-frequency stochastic process in its $12.5$~year dataset \citep{2020ApJ...905L..34A}, whose spectrum is common across many of the pulsars in its array. For a fiducial model of the GWB generated by a population of SMBHBs, the median characteristic strain amplitude at a frequency of $1/\mathrm{year}$ is $1.92\times10^{-15}$. Likewise, the PPTA \citep{2021ApJ...917L..19G} and EPTA \citep{2021MNRAS.508.4970C} have found similar processes with median strain amplitude $2.2\times10^{-15}$ and $2.95\times10^{-15}$, respectively. All of these results are statistically consistent with one another, and broadly in alignment with theoretical expectations of the GWB spectrum from SMBHBs \citep{2021MNRAS.502L..99M}. Furthermore, the IPTA has also recently announced the discovery of a similar stochastic process with amplitude $2.8\times10^{-15}$ \citep{2022MNRAS.tmp...73A}, where this result derives from the synthesis of older NANOGrav, PPTA, and EPTA datasets. However, none of these results exhibit significant inter-pulsar correlations, and as such these processes could still be noise or other systematic processes of non-GW origin. Nevertheless, it has been argued that the first milestone of GWB detection is the initial emergence of a common-spectrum process in PTA datasets, followed later by significant inter-pulsar correlations \citep{2021PhRvD.103f3027R,2021ApJ...911L..34P}. The latter measurement is the definitive evidence for a GWB, and requires that data from many pulsars be cross-correlated to infer the Hellings \& Downs signature. 

Modern PTA data-analysis frameworks model all signal and noise processes in the time domain. This is primarily for practical reasons. The signal of interest is a low-frequency stochastic process whose power-spectral density (PSD) is steeper than the PSD of a rectangular window function for existing pulsar-timing baselines; as such, a direct Fourier transform of the data without careful windowing would suffer from spectral leakage effects. Likewise, the observations are unevenly sampled, again rendering traditional Fourier transform techniques invalid (although Lomb-Scargle periodogram techniques could be used). The drawback of time-domain modeling is that all observations across all pulsars must be cross-correlated within a Gaussian likelihood function. Evaluating this likelihood requires the inversion of a dense covariance matrix, which scales with the number of timing observations as $\propto\mathcal{O}(N_\mathrm{obs}^3)$. Furthermore, with Bayesian techniques now the standard statistical approach, this likelihood function must be sampled at least $\gtrsim 10^6$ times across large parameter spaces in a global search over signal and noise parameters. As such, given that the total number of observations in recent NANOGrav data releases exceeds $10^5$ \citep{2021ApJS..252....4A}, even direct time-domain modeling is not tractable. 

The remedy to this has been rank-reduced approaches \citep{lha+13,vhv14,vhv15}, where random Gaussian processes are modeled in the time domain with basis design matrices that are much lower in dimensionality than the number of observations. The bottleneck of the likelihood remains a Cholesky covariance matrix inversion, but now of order the number of basis coefficients rather than the number of observations. The number of basis coefficients includes the number of timing ephemeris parameters, and the number of frequencies on which the GWB is modeled, which is small since we are targeting a low-frequency signal. Thus, rank-reduced modeling approaches enable PTA GWB searches to be carried out within reasonable periods of time ($\sim$~days to weeks). However, even this approach is becoming costly, and will struggle to cope in the future with the expansion of PTAs to include new pulsars and data e.g., from daily CHIME observations. Hence, in this paper we propose a new analysis scheme that parallelizes PTA data analysis over pulsars, and compresses pulsar data into sufficient statistics that can be arbitrarily and quickly combined in post-processing. The resulting approach is fast through the trivial parallelization of the analysis, and with a modularity that can cope with the computational demands of the ever-growing PTA data volume in the future. 

This paper is laid out as follows. We outline our approach for parallelization of Bayesian PTA data analysis in Sec.~\ref{sec:pta_like}. In Sec.~\ref{sec:results} we carry out a sequence of tests of this new approach, including assessing its performance for parameter estimation of the GWB's amplitude, and recovering the significance of a common-spectrum process and GWB-induced inter-pulsar correlations. We conclude in Sec.~\ref{sec:conclusions}, where we also remark on future applications of this method and further generalizations that we are exploring.

\section{Parallelizing the PTA Likelihood} \label{sec:pta_like}

GW searches with PTAs begin with data in the form of timing residuals, i.e., a best-fit deterministic timing ephemeris has already been constructed in terms of the pulsar period, spindown, binary parameters, etc., and the ephemeris-modeled TOAs then subtracted from the observed TOAs. These residuals encapsulate noise and any phenomena that were unmodeled at the stage of the original timing epehemeris construction. The latter group includes the influence of GW signals on the TOAs. A GWB will imprint spatio-temporal correlations in PTA data; these include long-timescale correlations in the residual time-series as a result of the expected low-frequency characteristic strain spectrum, as well as a distinctive pattern of inter-pulsar correlations known as the Hellings \& Downs curve \citep{1983ApJ...265L..39H}. This curve is the relevant correlation pattern imparted by an isotropic, stationary, and unpolarized Gaussian GWB on PTA data, and depends only on the angular separation between pulsars. 

\begin{figure}
    \includegraphics[width=\columnwidth]{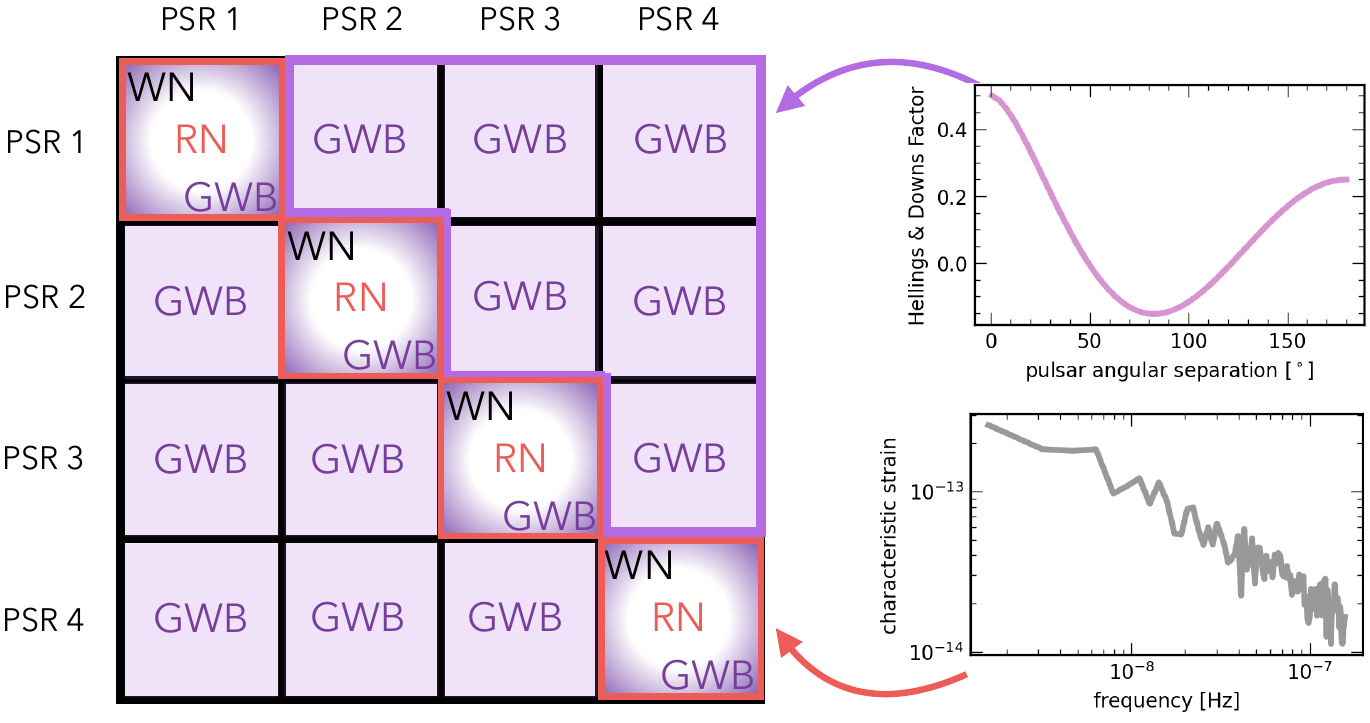}
    \caption{A schematic representation of the PTA data covariance matrix, where the pulsar autocorrelations are influenced by intrinsic white noise (typically instrumental), intrinsic red noise (typically low-frequency pulsar rotational instabilities), and a GWB signal. By contrast, the cross-correlations are influenced only by the GWB, and encode the sufficient evidence for a GW detection claim. On the other hand, the pulsar auto-correlations will dominate the spectral characterization of the GWB, since the cross-correlations are weaker.}
    \label{fig:pta_covariance_schematic}
\end{figure}

As a result, the entire PTA dataset can be modeled with a multi-variate Gaussian likelihood that has a dense time-domain covariance matrix. The elements of this matrix encode the temporal covariance expected between different timing residuals at different times, modulated by the corresponding Hellings \& Downs factor for the relevant angular separation between the pulsars. We write the PTA likelihood as 
\begin{equation}
    p(\boldsymbol{\delta t} | \boldsymbol{\eta}) = \frac{\exp\left( -\frac{1}{2}\boldsymbol{\delta t}^T C^{-1} \boldsymbol{\delta t}\right)}{\sqrt{\mathrm{det}(2\pi C)}},
\end{equation}
where $\boldsymbol{\delta t}$ is the vector of concatenated timing residuals across all pulsars in the PTA, and $C\equiv C(\boldsymbol{\eta})$ is the data covariance matrix with parameters $\boldsymbol{\eta}$. This covariance matrix can be described by
\begin{equation}
    \langle \boldsymbol{\delta t}_{ai} \boldsymbol{\delta t}_{bj}^T \rangle = C_{(ai)(bj)} = N_{a,(ij)}\delta_{ab} + C^\mathrm{red}_{a,(ij)}\delta_{ab} + \Gamma_{ab}C^\mathrm{GWB}_{(ij)},
\end{equation}
where $\delta_{(\cdot\cdot)}$ is the Kronecker delta function, $(a,b)$ index pulsars, $(i,j)$ index timing residuals, $N$ is a white noise covariance matrix in a given pulsar, $C^\mathrm{red}$ is the intrinsic red-noise covariance matrix in a given pulsar, $C^\mathrm{GWB}$ is the GWB covariance matrix, and $\Gamma_{ab}$ is the Hellings \& Downs cross-correlation factor between the relevant pair of pulsars, given by
\begin{equation}
    \Gamma_{ab} = \frac{3}{2}x_{ab}\ln x_{ab} - \frac{x_{ab}}{4} + \frac{1}{2} + \frac{\delta_{ab}}{2},
\end{equation}
and $x_{ab} = (1-\cos\theta_{ab})/2$ for pulsars with angular separation $\theta_{ab}$. A schematic for the structure of this PTA covariance matrix is shown in \autoref{fig:pta_covariance_schematic}.

Recent work \citep{2021PhRvD.103f3027R} has shown that the curation of existing PTA datasets has led to several high-quality pulsars being observed over long baselines, with the potential result that key pulsars now lie in the \textit{intermediate signal regime} of GWB detection; in \citet{2013CQGra..30v4015S} this is defined as the GWB signal exceeding the level of intrinsic pulsar and instrumental noise in lower frequencies in the power spectrum of pulsar timing residuals. An additional effect is that the expected GWB-induced cross-correlations are no greater than a half of the auto-correlation values (i.e., the Hellings \& Downs curve has a maximum of 0.5). As such, the significance of a common-spectrum process (i.e., information encoded in the auto-correlations) should currently be much greater than the significance of GWB-induced cross-correlations. This has been shown in simple analyses \citep{2021PhRvD.103f3027R}, as well as more realistic simulations that mirror the existing cadence and sensitivity of real datasets \citep{2021ApJ...911L..34P}. Thus, the working explanation for the emergence of strong common-spectrum processes in the datasets of all three long-baseline PTA collaborations \citep{2020ApJ...905L..34A,2021ApJ...917L..19G,2021MNRAS.508.4970C} is that this is an early indicator of an emerging GWB signal at low frequencies, to be followed several years later by the definitive GWB evidence in the form of significant cross-correlation measurement \citep{2021ApJ...911L..34P}.

With the information content of GWB auto-correlations currently swamping that of cross-correlations, the PTA covariance matrix can be approximated with a block-diagonal structure. While we can not use this auto-correlation information \textit{directly} to claim evidence of a GWB, we can use it for accurate, fast, and parallelizable spectral characterization. To see this, we now drop the cross-correlation blocks of the PTA covariance matrix, such that
\begin{align}
    p(\boldsymbol{\delta t} | \boldsymbol{\eta}) &\approx \frac{\exp\left( -\frac{1}{2}\sum_{a}\boldsymbol{\delta t}_a^T C_{aa}^{-1} \boldsymbol{\delta t}_a\right)}{\sqrt{\mathrm{det}(2\pi C)}}, \nonumber\\
    &= \prod_{a} \frac{\exp\left( -\frac{1}{2}\boldsymbol{\delta t}_a^T C_{aa}^{-1} \boldsymbol{\delta t}_a\right)}{\sqrt{\mathrm{det}(2\pi C_{aa})}}, \nonumber\\
    &= \prod_{a} p(\boldsymbol{\delta t}_a | \boldsymbol{\eta}).
\end{align}
We have now factorized the PTA likelihood into a product of per-pulsar likelihoods. This then allows independent, parallel analyses to be performed in each pulsar dataset, with these intermediate results then stitched together in post-processing to yield the full PTA result. 

In this paper we consider a GWB with the fiducial power-law characteristic strain spectrum from a population of GW-emitting SMBHBs, $h_c(f) = A_\mathrm{GWB}(f/1\,\mathrm{yr}^{-1})^{-2/3}$ \citep[e.g.,][]{2001astro.ph..8028P}, which yields a timing-residual power spectral density (PSD), $S_\mathrm{GWB}(f) = (A_\mathrm{GWB}^2/12\pi^2)(f/1\,\mathrm{yr}^{-1})^{-\gamma}\,\mathrm{yr}^3$, where $\gamma=13/3$. As such, in each independent pulsar dataset we sample intrinsic pulsar and instrumental noise, in addition to the amplitude, $A_\mathrm{CP}$, of a low-frequency common process that has a fixed PSD spectral-index of $\gamma=13/3$. This latter common process is a proxy for the GWB in an individual pulsar dataset. Since $A_\mathrm{CP}$ is the only common search parameter across the array, we can use its marginal posterior distribution from each pulsar as a sufficient statistic to recover the marginal posterior of $A_\mathrm{CP}$ for the entire array, i.e.,
\begin{align}
    p(A_\mathrm{CP} | \boldsymbol{\delta t}) &\propto p(A_\mathrm{CP}) \prod_{a} \frac{p(A_\mathrm{CP} | \boldsymbol{\delta t}_a)}{\bar{p}(A_\mathrm{CP})} ,
\end{align}
where $\bar{p}(A_\mathrm{CP})$ is the prior on the common-process amplitude applied during the analysis of each pulsar dataset, and $p(A_\mathrm{CP})$ is the prior on the final, desired PTA result. The practical details of how this Factorized Likelihood (FL) scheme is implemented are given in the next section, along with the means by which FL results can also accelerate the search for cross-correlation significance.  

\section{Results}\label{sec:results}

We test the efficacy of our new FL approach against the full PTA likelihood function using a suite of realistic PTA simulated datasets. The format of these simulations closely follow \citet{2021ApJ...911L..34P}. The PTA configuration corresponds to the $45$ pulsars from the NANOGrav 12.5~year dataset \citep{2021ApJS..252....4A} that were searched for the influence of an isotropic stochastic GWB \citep{2020ApJ...905L..34A}. Observational timestamps, TOA uncertainties, noise properties, and other metadata were derived from the real pulsar data, and then used to create synthetic noise-only data realizations. Rather than restrict to a maximum of $12.5$~years of data, we extend to a near-future scenario corresponding to $15$~years. This is achieved by drawing observational cadences and uncertainties from distributions of the final year of real data, then adding new observations to each pulsar until the maximum baseline of $15$~years is reached. No new pulsars are added to the array when expanding from $12.5$~years to $15$~years \footnote{This is a conservative choice for simulations, given that NANOGrav and other PTAs are constantly adding new pulsars. However, the addition of new pulsars is irrelevant here since we are only interested in creating a realistic set of simulations to assess the relative efficacy of the analysis approaches.}.

We then inject $100$ different realizations of a GWB signal into these noise-only datasets, generating a set of $100$ PTA datasets, each of which has $45$ pulsars. One of our key aims is to verify that the FL approach has equivalent Bayesian statistical coverage to the full PTA likelihood, i.e., the injected parameter values fall within the $p\%$ posterior credible region in $p\%$ of realizations. In Bayesian inference, this is only guaranteed when the injected parameter values are drawn from within the prior distribution that is subsequently used in the analysis. Thus, the spectral characteristics of the GWB used to generate each of the $100$ injections are drawn from  $\log_{10}A_\mathrm{GWB} \in U[-18,-14]$, where the spectral index is fixed at $\gamma_\mathrm{GWB}=13/3$. 

The basic analyses to be performed for all of the following tests are Bayesian per-pulsar noise characterization studies. Since we are ignoring inter-pulsar correlations, these analyses are performed entirely in parallel. The inference model of each corresponds to white noise fixed at the injected level, with intrinsic red noise characterized by a power-law PSD with search parameters $\log_{10}A_\mathrm{red}\in U[-20,-11]$ and $\gamma_\mathrm{red} \in U[0,7]$. We also model an additional power-law PSD process with fixed spectral index $\gamma_\mathrm{CP}=13/3$, and $\log_{10}A_\mathrm{CP} \in U[-18,-14]$; this process acts as a modeling proxy for the presence of the GWB in each pulsar. The result of analyzing each PTA dataset is a sequence of $45$ $\times$ $3$-dimensional MCMC chains, each iterated for at least $10^6$ steps, and subsequently thinned by a factor of $10$ (i.e., only every $10$th sample was used) to deliver $10^5$ usable samples. Each chain is post-processed into a histogram representation of the $1$-dimensional marginalized posterior distribution of $\log_{10}A_\mathrm{CP}$ in the respective pulsars. While there are various rules-of-thumb to deduce the optimal bin-width in such a histogram representation (e.g., Scott's rule \citep{scott1979optimal}), these usually assume that the underlying density is Gaussian. A more agnostic scheme would implement the Sheather-Jones algorithm \citep{sheather1991reliable}, which attempts to minimize the asymptotic mean integrated squared error between the estimator and the true underlying density. In practice, we used $N_\mathrm{bin}=100$ for most of our results, however this choice is varied in Sec.~\ref{sec:dropout}. The end result of analyzing all pulsars within a given PTA dataset is that the data has been compressed into histogram estimators of the marginal posterior of $\log_{10}A_\mathrm{CP}$. 

\subsection{Parameter estimation} \label{sec:parameter_estimation}

\begin{figure*}
    \includegraphics[width=\columnwidth]{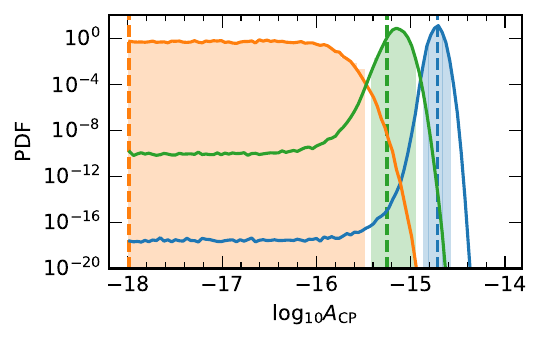}
    \includegraphics[width=\columnwidth]{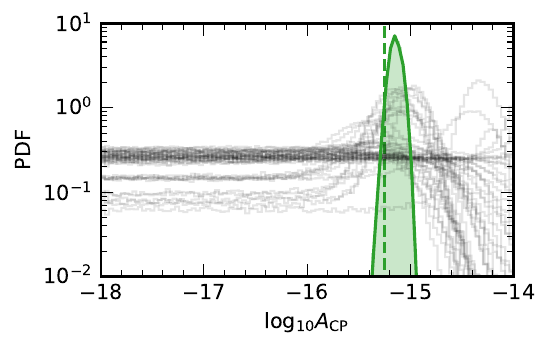}
    \caption{\textit{Left:} Example of common process amplitude posterior recovery for a weak, moderate, and strong signal injection. Faded regions are from MCMC sampling of the full PTA likelihood. Solid lines are parallelized reconstructions using the factorized likelihood technique. Dashed vertical lines are the injected values. \textit{Right:} The FL amplitude posterior recovery of the moderate signal from the left panel is shown again here, along with the common process amplitude posteriors from the individual pulsar analyses (faded grey histograms). These individual pulsar posteriors are the compressed representation of the pulsar data that are used for the final PTA recovery (green).}
    \label{fig:amplitude_recovery_exampls}
\end{figure*}

With our entire PTA datasets distilled down to a set of histogram probability density estimators of the common-spectrum process' amplitude, parameter estimation is remarkably simple. Given that the prior on this amplitude was uniform in $\log_{10}A_\mathrm{CP}$, all histogram estimators can be multiplied together and renormalized to deduce the equivalent full-PTA posterior of $\log_{10}A_\mathrm{CP}$. One pragmatic refinement is that occasionally some pulsars will lack MCMC-sample coverage over the same values as others, leading to multiplication of finite values by zeros in some bins. This can result in sharp posterior declines or poor posterior recovery at higher, data-informed values of the amplitude. To remedy this, we add a small value of $\epsilon=10^{-20}$ to all bins in all pulsar histograms. 

In the left panel of \autoref{fig:amplitude_recovery_exampls}, we use three simulations (corresponding to strong, moderate, and weak signal injections) to contrast the performance of our FL estimation of $\log_{10}A_\mathrm{CP}$ with the full PTA likelihood model. Not only is the FL estimate equivalent to what is deduced through sampling with the full PTA likelihood, it is also much less inhibited by sampling limitations in low density regions. It is able to recover the posterior density many orders of magnitude lower than what direct MCMC sampling with the PTA likelihood can achieve (compare the FL estimation as solid lines in \autoref{fig:amplitude_recovery_exampls} to direct MCMC sampling as shaded histograms), which results from sampling the tails of many individual pulsar posteriors that have less contrast than the full PTA posterior. We will see the enormous benefit of this for model selection in the next sub-section. The right panel of \autoref{fig:amplitude_recovery_exampls} illustrates the FL recovery of the moderate GWB signal injection in the PTA, alongside the recoveries of the amplitude of the proxy process in each individual pulsar.

\begin{figure}
    \includegraphics[width=\columnwidth]{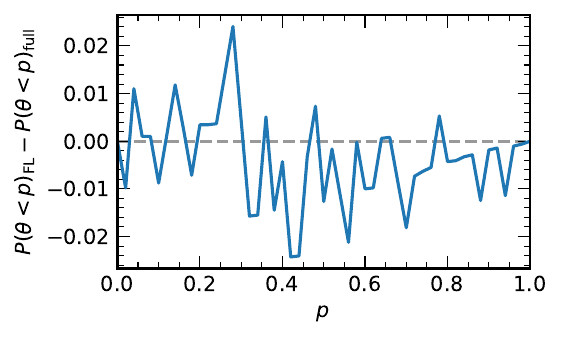}
    \caption{Bayesian $p-p$ comparison of amplitude recovery between the FL approach and the standard full PTA likelihood. The $y$-axis shows the difference that these two approaches give for the fraction of simulations in which the injected value lies within the $p\%$ credible region on the $x$-axis. The horizontal line is for equivalent efficacy of the approaches.}
    \label{fig:pp_plot}
\end{figure}

While the case studies shown in \autoref{fig:amplitude_recovery_exampls} are very promising, we now undertake a more rigorous assessment of the parameter estimation efficacy of the FL approach. The simulated dataset generation that was discussed earlier was specifically constructed to enable tests of Bayesian coverage through $p$$-$$p$ plots. By drawing injected signal parameters from within the analysis prior, the injected values should lie within the $p\%$ credible region in $p\%$ of simulations. Rather than judge the overall efficacy of the standard PTA data analysis pipeline, the scope of our analysis is to compare the relative performance of the FL approach to that achieved with full PTA likelihood. The results of this comparison are shown in \autoref{fig:pp_plot}, where the difference in the Bayesian coverage of each approach is shown as a function of credible interval. If the FL approach perfectly matched what an analysis with the full PTA likelihood could achieve then the difference should be zero at all credible intervals. While there is some mild scatter around zero, the resulting match is excellent, indicating that the FL approach effectively reproduces the efficacy of the standard PTA analysis pipeline.

\subsection{Model selection}

The parallelized parameter estimation discussed in the previous section facilitates a two-stage model selection process for PTA GWB searches, namely: $(1)$ emergence of a spectrally-common stochastic process across pulsars in the array, and $(2)$ detection of Hellings \& Downs inter-pulsar correlations, regarded as the definitive detection signature.

\subsubsection{Detecting a common process} \label{sec:cp_detection}

The relevant detection statistic for a spectrally-common stochastic process is the Bayes factor between models that include this process versus one in which it is absent (i.e., per-pulsar noise alone). In our case, where the GWB spectral shape is fixed to $\gamma_\mathrm{CP}=13/3$, the amplitude parameter acts as a signal ``switch'' that encompasses the noise-only model at the lower prior boundary ($\log_{10}A_\mathrm{CP}=-18$), and the signal$+$noise model for all other values. As such, computation of this Bayes factor can be addressed with the Savage-Dickey (SD) approximation, which depends only on the prior to posterior density ratio at the amplitude for which the signal is effectively zero. Hence
\begin{equation}
    \mathcal{B}_\mathrm{CP} = \frac{p(\log_{10}A_\mathrm{CP}=-18)}{p(\log_{10}A_\mathrm{CP}=-18|\{d_i\}_{N_p})}.
\end{equation}

\begin{figure*}
    \includegraphics[width=\columnwidth]{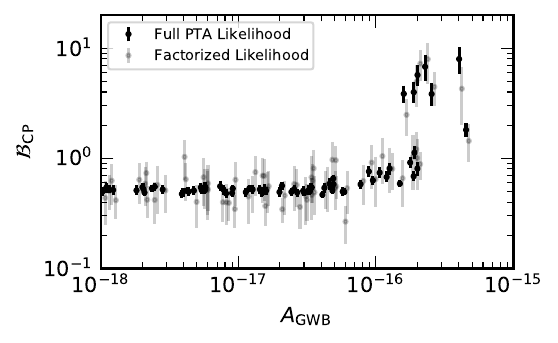}
    \includegraphics[width=\columnwidth]{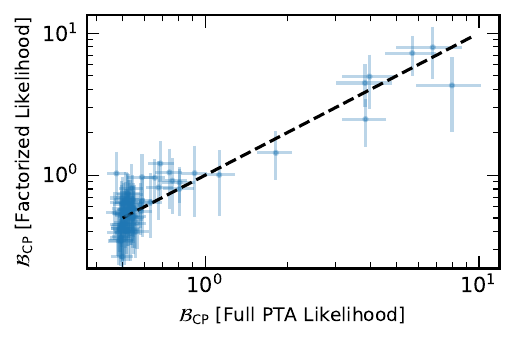}
    \caption{\textit{Left:} Bayes factors in favor of a common-spectrum process across all pulsars are shown as a function of the injected GWB amplitude. These are computed under the standard full PTA likelihood analysis, and using the parallelized FL analysis. The FL values are horizontally offset for ease of viewing. \textit{Right:} The amplitude dependence from the left panel is collapsed over to more directly compare the Bayes factor computed under each approach. The black dashed line is a zero-intercept unit-gradient relation to represent perfect agreement.}
    \label{fig:m2av1_compare}
\end{figure*}

We evaluate this for our suite of simulations, using both the full PTA model analysis and the parallelized FL analysis. In the full PTA model analysis, the posterior density is practically evaluated using $p(\log_{10}A_\mathrm{CP}=-18|\{d_i\}_{N_p})=f_b/\Delta_b$, where $f_b$ is the fraction of total MCMC samples contained within $-18 \leq \log_{10}A_\mathrm{CP}\leq-18+\Delta_b$, and the choice of $\Delta_b$ is varied over $100$ values between $0.01$ and $0.1$, with the results averaged. The bin width $\Delta_b$ is chosen to be small enough so as not to encounter any significant changes in the posterior density away from the value at the lower prior boundary. In the FL analysis, the normalized histogram estimator directly provides the posterior density. In both approaches, we estimate the uncertainty on the Bayes factors using bootstrap resampling. In the full PTA model analysis the single MCMC chain is resampled with replacement, while in the FL analysis each pulsar's MCMC chain is resampled with replacement before forming new histogram estimators that are multiplied across. This procedure is repeated $100$ times for each approach, generating distributions of Bayes factors that are summarized as median values with $68\%$ uncertainty regions. The results are shown in \autoref{fig:m2av1_compare}, where the left panel exhibits the growth of this common-process Bayes factor with GWB amplitude. The results deduced under each approach match well, as is shown more directly in the right panel where the amplitude dimension is collapsed over and the Bayes factors are directly plotted against one another. 

\subsubsection{Detecting inter-pulsar spatial correlations}

While evidence for a common-spectrum process across the array is a necessary condition for PTA GWB detection, it is not sufficient. It is possible that statistically-independent (yet spectrally-similar) low-frequency noise may arise in millisecond pulsars, which could appear as a common-spectrum process. Likewise, solar-system ephemeris or clock systematics could manifest as low-frequency processes in pulsar timing-residual data. As such, the only robust metric for GWB detection is via the measurement of Hellings \& Downs inter-pulsar timing-residual correlations, corresponding to the overlap reduction function for an isotropic stochastic background composed of tensor-transverse GW polarizations. The signature is predominantly quadrupolar in pulsar angular-separation, with power at higher multipoles but none at $l=0,1$. By contrast, intrinsic pulsar noise has zero expected cross-correlation with other distinct pulsars, while solar-system ephemeris and clock systematics are expected to produce dipolar and monopolar correlations, respectively \citep{2016MNRAS.455.4339T}.

The FL approach can be used to construct a statistic that measures the significance of cross-correlations, whether they be Hellings \& Downs or another template signature. This statistic is the optimal two-point correlation statistic for a GWB in PTA data---commonly referred to as the optimal statistic (OS)---and provides a signal-to-noise ratio (S/N) estimate for cross-correlations in PTA data \citep{2009PhRvD..79h4030A,2013CQGra..30v4015S,2015PhRvD..91d4048C,2018PhRvD..98d4003V}. The OS can be shown to derive from the linear term in a Taylor expansion of the PTA log-likelihood with respect to cross-correlations \citep{2009PhRvD..79h4030A,2013ApJ...769...63E,2015PhRvD..91d4048C}, and as such it is only applicable in the weak signal regime (where signal strength here refers to the relative amplitude of cross-correlated power versus auto-correlated power). However, an approach has been developed to improve the efficacy of the OS in intermediate signal scenarios, where it is marginalized over the joint posterior distribution of intrinsic pulsar noise and auto-correlated GWB power that has been computed from a full-PTA search for a common process \citep{2018PhRvD..98d4003V} \footnote{This Bayesian search for a common process, followed by computing the NMOS for evidence of cross-correlations, is still much computationally cheaper than a Bayesian search that models cross-correlations}. Both intrinsic noise and auto-correlated GWB power feature as individual pulsar weightings in the construction of the OS; by mapping the distribution of the statistic under the posterior spread of these weighting contributions, we propagate all sources of uncertainty in the noise estimation into the final OS distribution. This approach is referred to as the Noise Marginalized OS (NMOS). It is important that the calculation of the NMOS S/N use Monte Carlo draws from a joint posterior derived from a full PTA search, so that the covariance between individual pulsar red noise parameters and the common process (i.e., the auto-correlation term of the GWB) is accurately reflected. 

As shown previously in this paper, the FL approach gives an almost identical reproduction of the common red-noise posterior when compared to a conventional full-PTA analysis. Therefore we would expect it to give an equally faithful reproduction of NMOS results. However, to do so we must re-weight the posterior distributions from each of our separate pulsar analyses in order to achieve consistency with the FL-reconstructed common-process posterior. This is equivalent to using the FL-reconstructed marginal distribution of the common process amplitude as a weighting to piece together the full, joint PTA posterior distribution from our separate per-pulsar analyses. By doing so, the covariance between all red noise models across the PTA is properly incorporated just as in a conventional, full PTA Bayesian analysis.

Mathematically, this is achieved through re-weighting the conditional distribution of red noise parameters from an individual pulsar analysis, $p(\{A_\mathrm{red}, \gamma_\mathrm{red}\}_i|A_\mathrm{CP}, d_i)$, by the ratio of the FL-reconstructed common process posterior $p(A_\mathrm{CP}|\{d_i\}_{N_p})$ to the common-spectrum process posterior in that individual pulsar, $p(A_\mathrm{CP}|d_i)$: 
\begin{align}
    p(\{A_\mathrm{red}, \gamma_\mathrm{red}\}_i|A_\mathrm{CP},\{d_i\}_{N_p}) =&\,\, p(\{A_\mathrm{red}, \gamma_\mathrm{red}\}_i|A_\mathrm{CP},d_i) \nonumber\\ &\times\frac{p(A_\mathrm{CP}|\{d_i\}_{N_p})}{p(A_\mathrm{CP}|d_i)}.
\end{align}
The calculation of the NMOS is then as follows: $(i)$ a random draw is made from the FL posterior of $A_\mathrm{CP}$; $(ii)$ the re-weighted joint posterior of a pulsar's red noise parameters given this value of $A_\mathrm{CP}$ is constructed, and then used to draw values for the red noise parameters. This procedure ensures that the drawn samples capture the same covariance structure present in the posteriors from a full Bayesian analysis. Stage $(ii)$ is repeated for every pulsar in the PTA. The entire process is repeated $10^4$ times to give a distribution of OS values. Practically, we carry out this procedure by creating a three-dimensional empirical distribution for each pulsar from the posteriors for $(A_\mathrm{red}, \gamma_\mathrm{red}, A_\mathrm{CP})$. The 3D distribution is then collapsed to a 2D conditional distribution for $(A_\mathrm{red}, \gamma_\mathrm{red})$ for each value of $A_\mathrm{CP}$ drawn from the FL posterior. The red noise parameters used in the NMOS calculation are then drawn from each individual pulsar's 2D conditional red-noise distribution.

\begin{figure*}
    \includegraphics[width=\columnwidth]{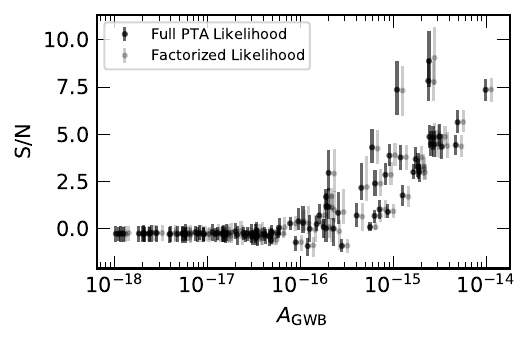}
    \includegraphics[width=\columnwidth]{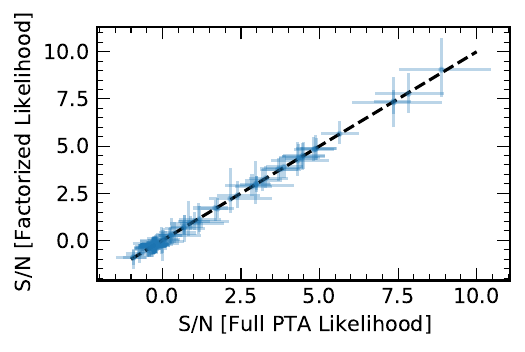}
    \caption{\textit{Left:} NMOS S/N values for Hellings \& Downs cross-correlations are shown as a function of the injected GWB amplitude. The conventional calculation involves marginalizing the OS values over noise and common-process samples generated from a Bayesian analysis with the full PTA likelihood. In the FL approach, noise values for each pulsar are drawn from re-weighted conditional distributions based on the FL-reconstruction of the common process amplitude, thereby imprinting the covariance structure of the full PTA on individual-pulsar posterior distributions. The FL values are horizontally offset for ease of viewing. \textit{Right:} The amplitude dependence from the left panel is collapsed over to more directly compare the S/N values from each approach. The black dashed line is a zero-intercept unit-gradient relation to represent perfect agreement.}
    \label{fig:nmos_compare}
\end{figure*}

\autoref{fig:nmos_compare} shows the results of applying this FL scheme for computing cross-correlation S/N values on our simulated datasets. The agreement with the conventional approach is very strong, showing virtually identical recovery in the median and spread of S/N values. The benefits of the FL approach are that this never requires a full PTA Bayesian analysis, pulsars do not need to be re-analyzed repeatedly, and this parallelized scheme easily allows for new pulsars to be analyzed and incorporated into the calculations.

\subsection{Cross-validation}\label{sec:dropout}

Parameter estimation and model selection are important inference tasks, but are not the only means by which confidence can be built in the detection of a signal. The model selection discussed so far has all been ``in sample'', i.e., all data is used to compute a figure of merit under different models, then the model with the largest figure of merit (in our case the Bayesian evidence) is favored. However, ``out of sample'' model selection can also be important in assessing how well a model is internally self-consistent among the data (see, e.g., \citet{2019MNRAS.487.3644W}). For example, leave-one-out cross-validation in PTA analysis consists of assessing a model's performance by conditioning it on all but one pulsar, then projecting it into the left-out pulsar to see how well it is supported. Essentially, we are testing the predictive power of a model that has been trained on a portion of the data then tested against another portion that has been held out.

In the PTA literature, leave-one-out cross-validation is referred to as ``dropout'', because each pulsar is dropped out sequentially to determine whether the model evidence for a common process increases or decreases as a result \citep{2019ApJ...880..116A,2020ApJ...905L..34A,2021ApJ...917L..19G}. Below we provide a mathematical framework for the dropout factor to serve as a guide for developing a FL approach. The dropout factor will be presented in full in a separate study \citep{dropout}.

Consider two hypotheses for an array of $N_p$ pulsars:
\begin{itemize}
    \item $\mathcal{H}_0$: intrinsic noise per pulsar, plus a common-spectrum process (with $13/3$ spectral index) in all $N_p$ pulsars. 
    \item $\mathcal{H}_1$: intrinsic noise per pulsar, plus a common-spectrum process (with $13/3$ spectral index) in $N_p-1$ pulsars, where pulsar $p$ is included in the analysis but does not respond to the common process.
\end{itemize}

The Bayesian evidence for $\mathcal{H}_0$ is
\begin{align}
    \mathcal{Z}_0 = p(\{d_i\}_{N_p}|\mathcal{H}_0) = &\int p(\{d_i\}_{N_p}|\{\vec{\theta}_i\}_{N_p},A_\mathrm{CP}) \times \nonumber\\
    &p(\{\vec{\theta}_i\}_{N_p})p(A_\mathrm{CP})\,\, d^{N_p}\vec{\theta} \,dA_\mathrm{CP}
\end{align}
where $p(\{d_i\}_{N_p}|\{\vec{\theta}_i\}_{N_p},A_\mathrm{CP})$ is the PTA likelihood for all $N_p$ pulsars, $p(\{\vec{\theta}_i\}_{N_p})$ is the prior for all intrinsic noise parameters, and $p(A_\mathrm{CP})$ is the prior for the common process' amplitude. As discussed in Sec.~\ref{sec:pta_like}, without any inter-pulsar correlations the PTA likelihood exactly factorizes into a product over pulsars. We can simply split this into the likelihood for $(N_p-1)$ pulsars (which is the likelihood for hypothesis $\mathcal{H}_1$) and the $p^\mathrm{th}$ pulsar.
\begin{align}
    p(\{d_i\}_{N_p}|\{\vec{\theta}_i\}_{N_p},A_\mathrm{CP}) =& \prod^{N_p}_{i=1} p(d_i|\vec{\theta}_i,A_\mathrm{CP}) \nonumber\\
    =& \,\,p(\{d_i\}_{N_p-1}|\{\vec{\theta}_i\}_{N_p-1},A_\mathrm{CP}) \nonumber\\
    &\times p(d_p|\vec{\theta}_p,A_\mathrm{CP}).
\end{align}
Therefore the evidence for $\mathcal{H}_1$ can be written as
\begin{align}
    \mathcal{Z}_1 = \int& p(\{d_i\}_{N_p-1}|\{\vec{\theta}_i\}_{N_p-1},A_\mathrm{CP})p(\{\vec{\theta}_i\}_{N_p-1})p(A_\mathrm{CP}) \nonumber\\
    &\times p(d_p|\vec{\theta}_p)p(\vec{\theta}_p)\,\, d^{N_p}\vec{\theta}\, dA_\mathrm{CP}
\end{align}
where we have used the fact that the prior on intrinsic noise parameters is also factorizable. Using Bayes' Theorem, we can recognize the first grouping of terms before the product symbol as $p(A_\mathrm{CP},\{\vec{\theta}_i\}_{N_p-1}|\{d_i\}_{N_p-1})\times \mathcal{Z}_*$, where $\mathcal{Z}_*$ is a normalization factor. Since this is a normalized posterior probability distribution, integrating over the common process amplitude and all intrinsic noise parameters of the $(N_p-1)$ pulsars considered in $\mathcal{H}_1$ simply yields $1$. The only remaining terms are an integral over the product of the likelihood and prior of intrinsic noise parameters in pulsar $p$, which is the Bayesian evidence for a noise-only model in this pulsar, $\mathcal{Z}_{p,1}$ (i.e., model $\mathcal{H}_1$). Finally, the evidence for model $\mathcal{H}_1$ is
\begin{align}
    \mathcal{Z}_1 &= \mathcal{Z}_* \int p(d_p|\vec{\theta}_p)p(\vec{\theta}_p)\,\, d\vec{\theta}_p\nonumber\\
    &= \mathcal{Z}_* \mathcal{Z}_{p,1}.
\end{align}

We now return to $\mathcal{Z}_0$ to factorize the likelihood, but remembering that the likelihood for pulsar $p$ depends on $A_\mathrm{CP}$ in model $\mathcal{H}_0$. We still get a grouping of terms that can be rewritten as $p(A_\mathrm{CP},\{\vec{\theta}_i\}_{N_p-1}|\{d_i\}_{N_p-1})\times \mathcal{Z}_*$, but one can only marginalize this distribution over the intrinsic noise parameters of which pulsar $p$ is independent. This gives the marginal distribution $p(A_\mathrm{CP}|\{\vec{\theta}_i\}_{N_p-1},\{d_i\}_{N_p-1})$ which we write compactly as $p(A|\mathcal{H}_1)$. The evidence for model $\mathcal{H}_0$ is then
\begin{align}
    \mathcal{Z}_0 &= \mathcal{Z}_* \int p(A_\mathrm{CP}|\mathcal{H}_1) p(d_p|\vec{\theta}_p,A_\mathrm{CP})p(\vec{\theta}_p)\,\, d\vec{\theta}_p dA_\mathrm{CP} \nonumber\\
    &= \mathcal{Z}_* \mathcal{Z}_{p,0} \int \frac{p(A_\mathrm{CP}|\mathcal{H}_1)}{p(A_\mathrm{CP})} p(A_\mathrm{CP},\vec{\theta}_p|d_p)\,\, d\vec{\theta}_p dA_\mathrm{CP},
\end{align}
where on the second line we have used Bayes' Theorem to express the calculation as an integral over the posterior distribution of pulsar $p$, and $\mathcal{Z}_{p,0}$ is the Bayesian evidence for a model that includes a common process in this pulsar (i.e., model $\mathcal{H}_0$). Finally, the ratio of model evidences $\mathcal{Z}_0 / \mathcal{Z}_1$---which is denoted as the dropout factor that assesses how well pulsar $p$ supports the presence of a common process found by the other $(N_p-1)$ pulsars---can be practically evaluated as
\begin{equation} \label{eq:FL_dropout}
    \mathrm{Dropout\,\, Factor} = \frac{\mathcal{Z}_0}{\mathcal{Z}_1} = \frac{\mathcal{Z}_{p,0}}{\mathcal{Z}_{p,1}} \left\langle \frac{p(A_\mathrm{CP}|\mathcal{H}_1)}{p(A_\mathrm{CP})}  \right\rangle_p
\end{equation}
where $\langle\rangle_p$ is an average over posterior samples from pulsar $p$. The right-hand side of \autoref{eq:FL_dropout} is easily computed using FL techniques, thereby avoiding the need to perform large numbers of PTA analyses and also rendering it trivial to assess the support for a common process in newly added pulsars. The one-dimensional marginalized posterior distribution $p(A_\mathrm{CP}|\mathcal{H}_1)$ can be deduced from an FL parameter estimation analysis in $(N_p-1)$ pulsars, similar to those shown in Sec.~\ref{sec:parameter_estimation}. The ratio of this posterior density to the known prior density of $A_\mathrm{CP}$ is then evaluated at, and averaged over, the posterior samples of the dropped-out pulsar $p$. The ratio of evidences on the right-hand side, $\mathcal{Z}_{p,0} / \mathcal{Z}_{p,1}$, gives the Bayes factor for a common red process in pulsar $p$, acting as a prior weighting for this common process in the calculation. It can easily be computed from the MCMC chain of each single-pulsar analysis using the Savage-Dickey technique. 

\begin{figure*}
    \includegraphics[width=\textwidth]{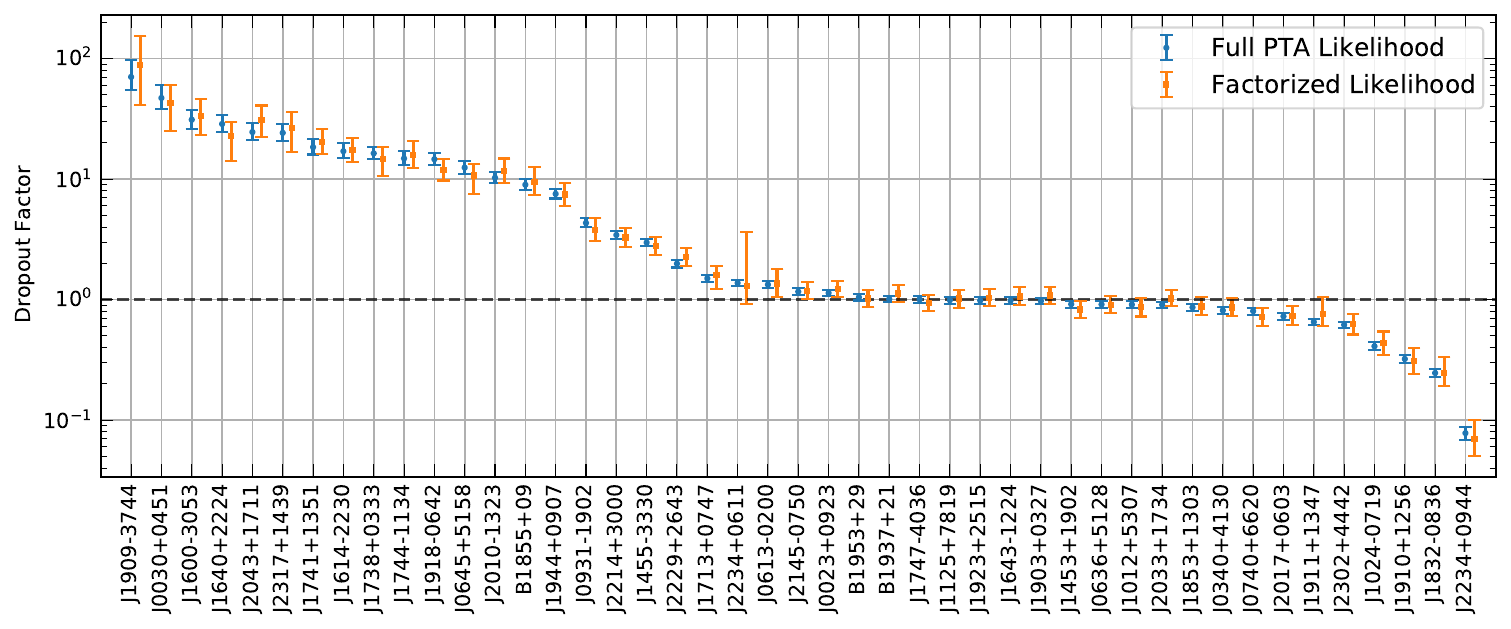}
    \caption{The dropout factor is the ratio of Bayesian evidences between a model that includes a common process in all pulsars versus one that leaves the common process out of one pulsar. This factor assesses the support that a given pulsar has for a common process that has been found by all other pulsars, acting as a technique for performing leave-one-out cross-validation. We compute the dropout factors for all pulsars in one of our simulated datasets that includes a GWB with $A_\mathrm{GWB}=1.91\times10^{-15}$. The conventional product-space MCMC sampling approach (blue) is compared to the new FL approach of \autoref{eq:FL_dropout} (orange), where we see excellent agreement. All uncertainties are derived through bootstrap resampling of MCMC chains.}
    \label{fig:dropout_compare}
\end{figure*}

We now compute the dropout factors for each pulsar in one of our simulated datasets, which contains an injected GWB signal with amplitude $A_\mathrm{GWB}=1.91\times 10^{-15}$ (broadly similar to the common process discovered in the NANOGrav 12.5yr dataset \citep{2020ApJ...905L..34A}). The conventional technique for computing the dropout factor is via a Bayesian product-space sampling analysis \citep[see, e.g.,][]{2018ApJ...859...47A,2020PhRvD.102h4039T}, where an MCMC chain samples a space of parameters that is concatenated over two models: one in which a common process is present in all pulsars, and another in which the common process is present in all but a selected pulsar. The transition between models is achieved by including an additional indexing parameter with different model behavior activated in different ranges, e.g., $n_\mathrm{model}\in[0,0.5]$ activates the model likelihood with a common process in all pulsars, whereas $n_\mathrm{model}\in[0.5,1]$ activates a likelihood with one pulsar dropped out.  
The ratio of samples for which the chain occupies one model range over another gives the odds ratio between the models, and thus (assuming equal prior odds between the models) the dropout factor. The uncertainty on this value is derived from $10^3$ bootstrap resamplings of the MCMC chain. 

Results for this conventional dropout approach are shown as blue points and uncertainties in \autoref{fig:dropout_compare}. We see that $\sim 1/2$ of pulsars support the common process found by the remainder of pulsars (i.e., dropout factors $>1$), $\sim 1/3$ are insensitive to the common process and yield dropout factors of $\sim 1$, and the rest disfavor the presence of a common process with dropout factors $<1$. This behavior is expected from the varying sensitivity of different pulsars; while it seems counter-intuitive that some pulsars disfavor a common process in a dataset that we know contains a GWB signal, one possible explanation is that we have ignored inter-pulsar correlations, which would otherwise exert greater control in driving agreement between the process amplitudes inferred from different pulsar subsets. This variant of the dropout factor---where inter-pulsar correlations are modeled---is still being developed and will be presented in \citet{dropout}.

We also analyze the dropout factors for this dataset using the FL-ready approach in \autoref{eq:FL_dropout}. As discussed earlier, a combination of the Savage-Dickey technique (for the first ratio of evidences) and FL posterior amplitude representation (for the integral under the posterior of a given pulsar) are needed. There are two metaparameters that we vary in this scheme. The first corresponds to the tolerance on the lowest acceptable number of MCMC samples in the lowest amplitude bin for the Savage-Dickey Bayes factor calculation of $\mathcal{Z}_p / \mathcal{Z}_{p,0}$. This influences the effective widths of the lowest amplitude bin whose posterior density features in the Bayes factor calculation discussed in Sec.~\ref{sec:cp_detection}. We vary this across five values: $10$, $25$, $50$, $75$, and $100$. The second metaparameter is the number of bins adopted for the histogram estimator of the common process posterior in the FL calculation. In our analyses thus far we found that $N_\mathrm{bin}=100$ was appropriate, however for this dropout factor analysis we allow it to be varied across $N_\mathrm{bin}=[10,25,50,75,100]$. Thus the dropout factor calculation for each pulsar was performed with $25$ distinct metaparameter combinations, and for each of those combinations the MCMC chains of all pulsars were resampled with replacement $1000$ times to generate $2.5\times10^4$ bootstrapped dropout factor estimates. These results are summarized as orange median values with $68\%$ uncertainty regions in \autoref{fig:dropout_compare}. The agreement with the conventional dropout approach is excellent. This FL dropout estimation approach is modular, all calculations are carried out in the MCMC post-processing stage, and new pulsars can be trivially incorporated. 

\section{Conclusions} \label{sec:conclusions}

We have introduced a new technique for gravitational-wave background (GWB) searches and characterization with pulsar-timing arrays (PTAs). This technique is parallelized over separate pulsar datasets, making it trivial to expand the array of searched pulsars without requiring exhaustive and repetitive re-analysis of the entire PTA dataset. By leveraging the fact that spectral information about the GWB is---and may remain so even in the post-detection era---dominated by PTA auto-correlations, we can collapse the PTA data covariance matrix to a block-diagonal structure. This is equivalent to approximating the PTA likelihood as a product over individual pulsar likelihoods, each of which includes a stochastic process with a common spectrum across the array. By doing so, each pulsar dataset can be analyzed in parallel, then reduced to a set of sufficient statistical measures for the spectrum of the GWB, which are stitched together in post-processing. We call this the Factorized Likelihood (FL) technique.

We have considered the case where the GWB is represented by a power-law characteristic strain spectrum, and equivalently a power-law power spectral density in the pulsar timing residuals. This power-law has a fiducial spectral index based on the theoretical expectations for the GWB from a population of inspiraling supermassive binary black holes. These supermassive binary black holes are expected to synthesize the dominant GWB signal in the PTA sensitivity band, in excess of other potential cosmological signals (e.g., primordial GWBs, and backgrounds from cosmological phase transitions or cosmic strings). As such, we can fix the spectral index and vary only the amplitude of the GWB in our searches. Therefore, in the analysis of each pulsar, we model noise associated with intrinsic pulsar and instrumental sources (with corresponding parameters), plus a stochastic process characterized by a fixed-index power-law, whose amplitude is searched over. This latter process is a proxy for the GWB in each individual pulsar's analysis.

Upon deriving the marginal posterior distributions of the GWB amplitude from each pulsar, we process these into histogram estimators that act as sufficient statistics for further analysis. All of the information about the GWB that was originally encoded in the PTA auto-correlations is now distilled into these histograms, which are de facto compressed representations of the PTA data. We found that the PTA GWB amplitude recovery using this FL approach was virtually indistinguishable from the standard PTA likelihood approach, as was the calculation of the Bayes factor for the presence of a common-spectrum process across the PTA. Furthermore, while this approach discards the PTA cross-correlations, it does facilitate the rapid and modular calculation of the PTA cross-correlation S/N, which otherwise requires an additional Bayesian PTA analysis to calibrate noise weightings. The S/N values derived from FL noise weightings versus a full Bayesian PTA analysis match incredibly well. Hence, the FL technique enables fast, parallelizable recovery of the amplitude and cross-correlation significance of a GWB in PTA data, thereby circumventing many of the sampling and computational limitations that PTAs will encounter as data volume grows. This technique has already seen broad uptake within the PTA community for analyzing NANOGrav \citep{2020ApJ...905L..34A}, PPTA \citep{2021ApJ...917L..19G}, EPTA \citep{2021MNRAS.508.4970C}, and IPTA \citep{2022MNRAS.tmp...73A} flagship datasets and other studies \citep{2022arXiv220110657J} (having been developed by the lead author), but has lacked a formal methodology until now. 

There are several further applications that are straightforward when tackled with the FL technique but which would otherwise be challenging using conventional sampling with the PTA likelihood. For example, in almost all PTA analyses the white-noise characteristics of each pulsar are held fixed at values previously found in single-pulsar noise characterization. The reasons for this are that the likelihood calculation is slowed down when sampling white-noise characteristics, and there can be many white-noise parameters per pulsar that lead to a greatly expanded parameter dimensionality. However since the FL technique is parallelized over pulsars, likelihood speed and sampling efficiency remain well within reasonable levels, thereby allowing the uncertainties in these white-noise characteristics to be propagated into the final GWB results. An extension of this concept is that bespoke noise models that account for e.g., additional chromatic influences, can be constructed for each pulsar in parallel, then trivially combined in post-processing to deliver PTA GWB constraints. Performing this same kind of analysis in a full PTA likelihood analysis is fraught with speed and sampling limitations.

The implementation of the FL technique introduced here does have some caveats that we plan to generalize in forthcoming work. Most important is that the spectrum of the GWB is assumed to be a power-law with a fixed spectral index, and parametrized only by its amplitude. Alternatively, we can model the GWB's spectrum in each pulsar with a free parameter in each frequency bin. Provided that the recovered joint posterior of this Bayesian periodogram from each pulsar can be represented with high fidelity (using e.g., optimized kernel density estimators), these then act as sufficient statistics for more generalized PTA spectral characterization of the GWB beyond the power-law assumption. With an optimally-tuned MCMC algorithm to sample the per-frequency spectral posteriors of each pulsar, this approach should be as fast and trivially parallelizable as the techniques presented in this paper. No significant additional computational time is required beyond the single-pulsar noise analyses that are already performed. This approach is currently being developed. The farthest we could push this FL concept would be to also recover the Fourier coefficients of the modeled GWB signal in each pulsar. By retaining phase information from the pulsar time-series in the form of the Fourier coefficients, we would regain the ability to perform Bayesian cross-correlation analyses, yet with a highly compressed representation of the original data. The end goal is that the PTA GWB analysis framework would be future-proofed against the torrent of data from ongoing observation campaigns as well as newly discovered pulsars.

\begin{acknowledgments}
We thank our colleagues in NANOGrav and the International Pulsar Timing Array for fruitful discussions and feedback during the development of this technique. We are particularly grateful to Michele Vallisneri for reviewing a draft of our manuscript, and to Xavier Siemens, Sarah Vigeland, and Jeff Hazboun for invaluable insight throughout this project. SRT acknowledges support from NSF AST-200793, PHY-2020265, PHY-2146016, and a Vanderbilt University College of Arts \& Science Dean's Faculty Fellowship. This work utilized resources from the University of Colorado Boulder Research Computing Group, which is supported by the National Science Foundation (awards ACI-1532235 and ACI-1532236), the University of Colorado Boulder, and Colorado State University. NP acknowledges support from the Vanderbilt Initiative for Data-intensive Astrophysics (VIDA). This work was conducted in part using the resources of the Advanced Computing Center for Research and Education (ACCRE) at Vanderbilt University, Nashville, TN.
\end{acknowledgments}

\bibliography{bib}

\end{document}